# A New Generation of Energy-Economy Modeling at the U.S. Energy Information Administration

J.F. DeCarolis, S. Siddiqui, A. LaRose, J. Woollacott, C. Marcy, C. Namovicz, J. Turnure, K. Dyl, A. Kahan, J. Diefenderfer, N. Vincent, B. Cultice, A. Heisey

All authors are affiliated with the U.S. Energy Information Administration

## Abstract

Given the rapid pace of energy system development, the time has come to reimagine the U.S. Government's capability to model the long-term evolution of the domestic and global energy system. As a primary custodian of these capabilities, the U.S. Energy Information Administration (EIA) is embarking on the development of a long-term, modular, flexible, transparent, and robust modeling framework that can capture the key dynamics driving the energy system and economy under a wide range of future scenarios. This new capability will leverage the current state of the art in modeling to produce critical insight for researchers, decision makers, and the public. We describe the evolving demands on energy-economy modeling, the capacity and limitations of existing models, and the key features we see as necessary for addressing these demands in our new framework, which is under active development.

## Table of Contents

1. Introduction

Energy models need to evolve to reflect the accelerating changes in the scale, scope, and complexity of the energy systems they represent. Both supply and demand of energy are changing atop an evolving technology landscape, policy environment, and physical climate. Energy markets are becoming more interconnected through continued globalization, and there is growing interest in assessing the full range of impacts from local to global.

U.S. energy supply is evolving rapidly. Over the past two decades primary energy production from fossil fuels has shifted from coal to natural gas and crude oil, and, in electric power, renewable supply has shifted from hydroelectric to wind and solar generation (U.S. EIA, 2024a). The economics and power grid impacts of wind and solar depend heavily on where and when they are available (Mai, 2018a; Cole, 2017), which require electricity models to consider the available resources and grid conditions at greater spatial (Krishnan, 2016) and temporal (Marcy, 2022) resolutions. In addition, there are many novel technology pathways emerging, including energy storage (Bistline, 2020), advanced nuclear (Bistline, 2023), hydrogen (Beagle et al. 2023), direct air capture (Realmonte et al., 2019), and synthetic fuels (König et al., 2015), whose deployment depends on many uncertain factors, including the degree of technology innovation, policy incentives, and public support. Furthermore, modeling boundaries are being pushed beyond the internal dynamics of the energy system to explicitly consider a broader scope; for example, supply chains and manufacturing infrastructure needed for the logistics and costs of these novel technology pathways (Moerenhout et al., 2023). Even the physical science of climate change is creating feedback loops into the energy system, which need to be accounted for (Craig et al., 2022).

Energy demand is also evolving in ways that require greater consideration in models. For example, end-use electrification, ranging from electric vehicles and household appliances to industrial-scale electric arc furnaces, could gradually reshape daily and seasonal load patterns that also affect the engineering and economics of electricity supply (Mai, 2018b). Furthermore, representing the adoption of these technologies endogenously increases model complexity (Mohseni et al 2022). End-use technology adoption and usage vary over socio-economic factors like income, age, and household composition (Kuang, 2023). Technology-specific consumer preferences also play a key role in determining technology adoption; for example, electric vehicle adoption depends on factors such as range anxiety, availability of charging infrastructure, and household travel patterns (Bunch et al., 2015), while the rapid market adoption of light-emitting diode (LED) light bulbs over their compact florescent (CFL) competitors was driven by consumer perception of performance and quality (Kelly, 2016).

Regional energy systems are interconnected and evolving under different mixes of drivers and constraints, which reflect global resource and economic realities. The scale of investment, expanding geographic and technological scope, and the increasing complexity of systems and supply chains suggests the rapid evolution of the energy system may have economic implications well beyond the energy system itself. Ongoing changes to the energy system are likely to have an impact on consumption, investment, government finance, trade, and employment, requiring modeling of integrated macroeconomic dynamics across all sectors.

Relevant macroeconomic and international factors include global energy prices, international policy, trade patterns, and geopolitical tensions. Most current energy system models for individual countries tend to make exogenous assumptions about macroeconomic and international interactions, but these factors are becoming increasingly intertwined with activity in the energy system. For example, many governments have targeted a transition to net zero greenhouse gas emissions over the next few decades (CRS, 2021). Independent estimates of the total cost associated with such a transition are as high as several trillion dollars per year over the next decade, about double current investment levels (BNEF, 2023). While EIA does not propose new policy, we must be able to model legally binding domestic and international laws and regulations.

Finally, there is deep uncertainty about these factors and the future of energy-economy dynamics more broadly. It has never been possible to model energy systems over the medium-to-long term with a high degree of precision (Huntington et al., 1982), but modeling efforts continue to produce useful insights about future energy systems that should be conditioned on the explicit consideration of uncertainty (NRC, 1992).

The time has come to reimagine the U.S. Government's long-term modeling capabilities to address these myriad issues. As a primary custodian of these capabilities, the U.S. Energy Information Administration (EIA) is embarking on the development of a modular, flexible, transparent, and robust modeling framework. Capturing the full suite of dynamics described above requires the ability to flexibly adjust model boundaries and resolutions across spatial, temporal, sectoral, and demographic dimensions. The modeled energy system should also remain integrated within a coherent economic framework that connects to the energy system across these dimensions. In the following sections, we define the core structure of an energy model and then we examine current modeling capabilities across the international modeling community, including limitations associated existing models, as well as key features associated with our next generation modeling framework.

## 2. Current Modeling Capabilities

Translating complex, real-world dynamics into a computational model requires breaking down the overarching challenge into manageable subsets. Here we characterize the general process of energy modeling with representative layers (Figure 1). Modern energy modeling involves developing both data (left column) and the model structure (right column) before models are instantiated and outputs are produced.

On the right side of Figure 1, under model structure, the top blue layer represents the governing dynamics, which denote the core modeling paradigm(s) used to drive systems change over time given a specified set of model input data. Examples include cost minimization, consumer choice, and economic equilibrium. Governing dynamics also include how policy, technology innovation, and economic growth drive changes over time. No matter how sophisticated the governing dynamics, they will always represent a simplification of the complex, real world dynamics at play.

The second layer (Figure 1, right) represents the system of mathematical equations that express the chosen governing dynamics. There may be multiple ways to express a given governing dynamic mathematically. These equations can often be expressed concisely in mathematical notation and are typically included in the model documentation. The third layer represents the code implementation, whereby the mathematical equations are implemented in a particular programming language. The choices among governing dynamics, mathematical equations, and programming language all have significant impacts on the model's computational performance and analytic strengths and weaknesses.

Energy system models are data intensive (Figure 1, left); multiple layers of data processing (shown in orange) are required to transform raw data into formatted model input data. The model code and input data are instantiated to conduct a particular model run (top gray layer). This step is handled automatically with software. Finally, the bottom layer represents the model output, which is often visualized in figures and tables. The model outputs with associated scenario specifications are the primary way that most observers draw insights from the modeling exercise.

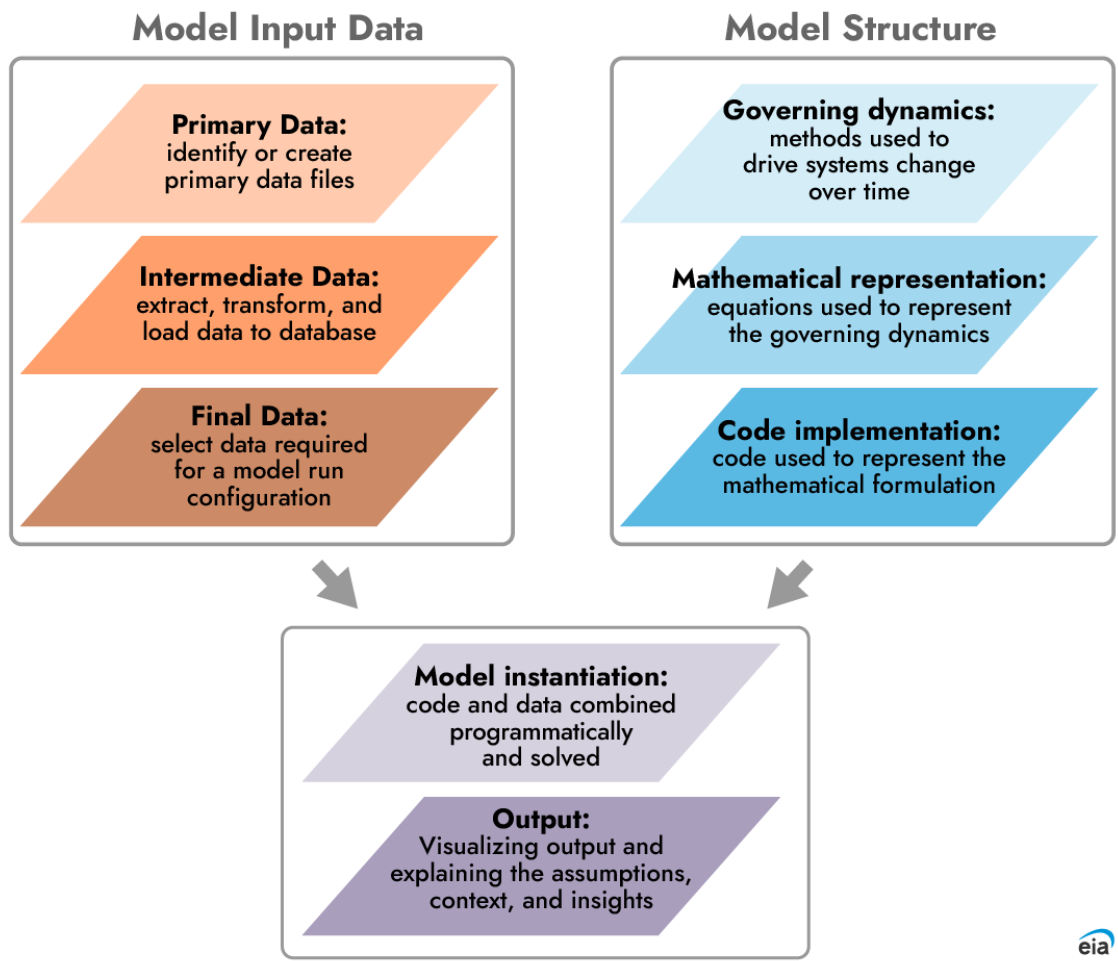

**Figure 1:** Representative layers associated with the development and application of an energy model. There are parallel layers pertaining to the development of model input data (left) and model structure (right). The data and model layers merge when a specific model run is instantiated and solved. In each of these layers, there is an opportunity to design EIA's next generation modeling framework to better align with our goals for modularity, flexibility, transparency, and robustness to best capture modern energy system dynamics.

## 2.1. Limitations of linear optimization models

Surveying the international energy modeling community (Openmod, 2024; ETSAP, 2024; Lopion et al., 2018; Pfenninger et al., 2014; Garguilo and O Gallachoir, 2013) we find that long-term energy models generally reflect a dominant governing dynamic and try to fit the entire system representation under it. For example, energy system optimization models, which are frequently used to develop long-term projections—Pfenninger et al. (2014), refer to them as the "backbone of energy systems modeling" — rely on cost minimization or welfare maximization dynamics to identify an energy system response to changing input costs, technologies, and demand over time and throughout the system. These models employ *linear optimization*, which is computationally tractable but simplifies the representation of key nonlinear dynamics to linear effects. While the use of linear optimization represents a convenient methodology, it is challenging to maintain the flexibility to introduce new governing dynamics without compromising the model's overall mathematical consistency and computational tractability. Sticking with a single governing dynamic for computational performance or other benefits can end up constraining model development options and precludes more faithful representations of certain real-world dynamics.

Energy system optimization models inherently take a prescriptive approach by finding the most efficient ways to evolve the energy system over time. An implicit assumption of these models is that there is an omniscient social planner who can optimally allocate resources among producers and consumers. Such an approach is particularly useful when examining ways to meet a long-term objective, like achieving net zero greenhouse gas emissions. In many circumstances, however, the goal is to anticipate realistic energy system responses to assumptions about the future rather than finding the most cost-effective pathways. Trutnevyte (2016) finds that projections based exclusively on optimization-based modeling approaches can deviate from real world outcomes in significant ways in part because optimization-based approaches do not account for the complexity and heterogeneity in decision making processes across the energy system.

The real-world heterogeneity in decision making is most lacking in the end-use sectors (i.e., residential, commercial, industrial, and transportation), where collective decisions can stray farthest from cost minimizing behavior (DeCarolis et al., 2017). Some energy system optimization models focus on energy supply and make fully exogenous assumptions in the demand sectors about stock turnover, future market shares, and end-use demands. Others represent the demand sectors endogenously through the optimal selection of demand technologies (e.g., vehicle, space heating, and lighting technologies) to meet elastic demands that are responsive to price. Because linear optimization often leads to "all-or-nothing" results whereby the cheapest options dominate the market, some models employ hurdle rates, growth rate limits, or market share constraints to prevent unrealistic shifts in technology choice, but which do not have a strong empirical basis (DeCarolis et al., 2017). Decision-making dynamics in the end-use sectors, while challenging to model, can have critical aggregate impacts on system outcomes.

A few modeling efforts go further by trying to introduce governing dynamics with a nonlinear formulation. For example, in the residential sector, responses to a change in electricity and fuel prices have substitution effects on technologies that are chosen, often captured by elasticities that have nonlinear functional formulations (e.g., Patankar et al., 2022). Other modeling efforts have embedded a nonlinear consumer choice formulation into an energy system optimization model (e.g., Ramea et al., 2018).

At EIA, the goal of our long-term outlooks is not to provide roadmap to achieve a desired outcome, but to assess how energy systems might realistically change under different assumptions about the future. Thus, our ability to experiment with governing dynamics that can better capture decision making dynamics is crucially important.

**2.2. Building on the National Energy Modeling System (NEMS)**

The National Energy Modeling System (NEMS) provides the projections published in EIA's *Annual Energy Outlook*. The current NEMS structure endogenously models demand with least-cost supply in independent, but connected, modules (U.S. EIA, 2023). NEMS incorporates some aspects of human decision making into relevant energy systems optimization, and mathematically represents policy and economic decisions within these modules. Further, the governing dynamics in one module are not compromised by the structure of the other modules, allowing for the most appropriate modeling representation in each sector. NEMS modelers can thus choose the best governing dynamics for each module. NEMS is sufficiently modular, but as explained below, there are opportunities to improve in terms of the flexibility, transparency, and robustness required in our long-term modeling program.

In terms of transparency, while EIA has made NEMS open source and publicly available, the current codebase remains difficult to understand and use.  NEMS has been under continual development for 30 years, which has led to a diverse mix of workflows and programming languages that make the model increasingly difficult to update, maintain, and execute in a secure, modern computing environment. Legacy code is also difficult to update. These very practical challenges make it difficult to overcome the technical limitations.

In terms of flexibility, a critical technical restriction of NEMS it can only reach equilibrium across the modules using the Gauss-Siedel algorithm (EIA, 2020), which iterates through each module, exchanging price and quantity information after each module is solved. A feature of the Gauss-Siedel algorithm is that it does not explicitly consider the internal mathematical structure of each module.  The existing Gauss-Seidel iteration results in long solve times and makes the total NEMS runtime heavily influenced by the slowest module. This feature also makes it very difficult to add new model innovations, as runtime concerns can drive decisions about model development. For example, increasing the spatiotemporal resolution in a single module could have a disproportionate impact on computational performance for the whole model given the need to iterate through all modules.

As discussed in more detail below, an alternative to Gauss-Seidel is to aggregate the mathematical structures in each module into a common mathematical formulation by an equivalence relation. This approach requires that all modules be programmed in the same language so that the full model can be instantiated all at once. Doing so in NEMS would require enormous effort to ensure that all existing modules are implemented in the same language. In addition, simply reimplementing the existing modules in a modern language limits our opportunity to think more broadly about model capabilities. In terms of Figure 1, the exercise would mostly be focused on the code implementation layer. Consideration of the governing dynamics and mathematical formulation would be anchored to our current formulation, depriving ourselves of the opportunity to freely design a model structure and functionality that ensures the efficient translation of information from one layer to another.

Finally, in terms of robustness, modeling sensitivity and uncertainty in NEMS can only be done by running different scenarios. Again, given the current NEMS workflow, scenarios can be difficult to setup and run individually. Thus, we would need to develop new capabilities to iterate the model to address model sensitivity and uncertainty.

## 3. Key Features of EIA's Next Generation Modeling

In 2022, we launched EIA's "Project BlueSky" to develop a new energy modeling framework. The name serves as a constant reminder that we need to take a step back from our existing models and think in a more expansive manner about the required capabilities of an EIA next generation model. At EIA, our long-term domestic and international outlooks must continue to focus on projecting realistic outcomes under a given set of scenario assumptions, accounting for rapid technological innovation, new policy implementation, changing consumer preferences, shifting trade patterns, and the real-world friction associated with the adoption of novel or risky technology.

There is no single formula or model configuration that will work in all circumstances. Using highly detailed versions of each module in all circumstances will quickly result in computational challenges, making it hard to tune the model to answer new questions as well as iterate the model under different assumptions to quantify uncertainty. Instead, the solution is to develop a modular and flexible framework that can be adjusted to address the question at hand (DeCarolis et al. 2017). High-level policy discussions and geopolitical events can materialize rapidly, and it is important to be able to quickly inform those discussions with independent and rigorous model-based analysis. This section explains how EIA's next generation model will be designed to be modular, flexible, transparent, and robust in its assessment of uncertainty related to the energy system and broader economy.

### 3.1. Modularity

The need to better capture different decision-making frameworks suggests a modular model structure, similar to NEMS. Unlike most existing models which incorporate a single dominant governing dynamic such as linear optimization (Figure 1), the next generation model will allow the governing dynamics to vary across different parts of the energy system and broader economy represented in each module (Figure 2). The modular structure provides a convenient means to test different approaches, isolate analysis to a particular part of the energy system or economy and vary the level of complexity by module depending on the question being addressed.

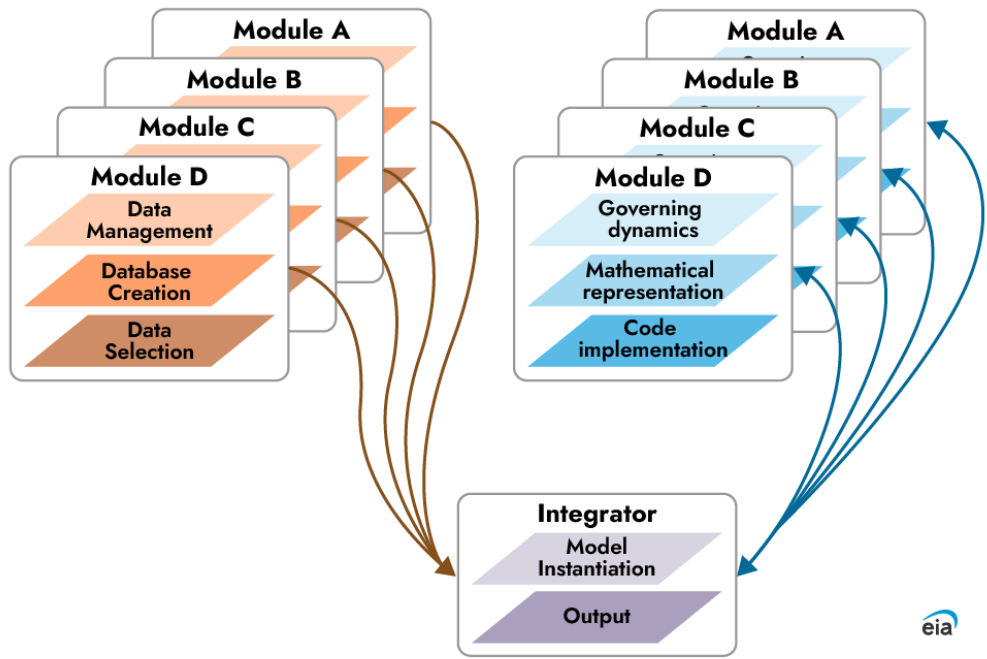

**Figure 2.** Conceptual framework for EIA's next generation model. Similar to NEMS, the governing dynamics, mathematical representation, and code implementation within the next generation model can vary by module. However, as described below, implementation of each module in the same programming language offers more computationally efficient ways to integrate modules and solve the model.

We draw inspiration from climate modelers, who have developed modular frameworks with component modules that can be run with varying levels of detail. Held (2005) notes that climate models are used both for *simulation*, where the model captures as many of the underlying dynamics as possible, and *understanding*, where the model is distilled into a version with only the core dynamics. Held (2005) recommends a model hierarchy on which to base our understanding, allowing modelers to evaluate how the model dynamics change as key model features are added or removed. For example, the National Center for Atmospheric Research (NCAR) Community Earth System Model (CESM) includes two simple atmospheric models to aid such hierarchical testing: "dynamical core" and "aquaplanet" (Polvani et al., 2017).

This approach similarly applies to energy system models. Like the physical earth system, the real-world complexity of energy systems is such that significant structural uncertainties in the model will always exist, which motivates testing different governing dynamics and using them to help bound model outputs. The key is to become proficient at understanding how aggregation or disaggregation relate to information loss in each specific element of the system being modeled and make thoughtful decisions about model structure on that basis. To be most effective, modelers must consider how well a given structure and set of assumptions can address the types of questions asked of a model while being mindful of how different model configurations may impact the insights drawn from the model.

A single model configuration is not suitable for all questions one might ask of a model – one size does not fit all. For example, addressing questions focused on aggregate outcomes from the energy system, such as market prices or emissions, may require endogenous interactions across all sectors, but may not require the most detailed resolutions available in all those modules. In other cases, questions may focus on the dynamics within a particular sector or pathway within the energy system, where it could be advantageous to run the model with maximal detail only in the portion of the model under investigation or even just run a standalone version of that module. Finally, some questions involving deep uncertainty about the future may require simplified modules that can be iterated many times to quantify uncertainty (e.g., Monte Carlo simulation). Or, they should have the ability to approximate sensitivity of model outputs to inputs in an efficient manner. A reduced-form representation can often be an appropriate portrayal of an interaction that extends outside the energy system, such as a market for carbon offsets or a complementary product market for an energy product.

Modularity has been a strength of NEMS from its earliest days. A National Research Council (NRC) report from 1992 focused on NEMS design recommended EIA develop "a set of modules linked together by a simple control module... so that they can be run separately, all together, or in combinations, depending on the analytic need." (NAS, 1992). The report further noted that the modular approach would enable "the substitution of alternative modules embodying different conceptual structures, theories, empirical representations, and databases," which could be used to assess structural uncertainties in the model (ibid.) EIA modelers developed a modular structure for NEMS but not interchangeable modules. EIA's next generation model offers an opportunity to develop this valuable functionality, enabling modelers to more readily test different versions of the same module with varying levels of complexity and test new approaches to support continued model innovation.

While the modular structure offers the many advantages discussed above, it does present computational challenges. To develop a self-consistent representation of the whole energy system with such a modular structure, each sector needs to exchange prices, quantities, and other sector-specific information for energy and other commodities to achieve an equilibrium.

One way to implement equilibrium in code is to use an iterative approach, where individual modules are run in an efficient order to exchange information until an equilibrium is achieved. As described above, NEMS uses Gauss-Seidel iteration, which mirrors the procedure for solving a system of linear equations. The advantage of such an iterative method is that if the iterations converge, we are guaranteed to achieve an equilibrium. It also can be used with modules that are programmed in different languages. As noted above, the disadvantage is that iterative procedures for achieving equilibrium are computationally inefficient because they do not take advantage of the known mathematical formulation of the individual modules. They are also not guaranteed to converge, and since the mathematical representations of each problem are different, it's very difficult to provide a theoretical foundation that guarantees convergence.

A second way to implement equilibrium is to integrate the modules into one mathematical structure that can contain a potentially diverse set of governing dynamics. Theoretically, if an efficient mathematical structure can be found, these methods are more computationally efficient than iterative methods of achieving equilibrium. For example, in solving formulations with multiple players, iterative methods such as Gauss-Seidel (also referred to as diagonalization in the literature) tend to converge slower than combining problems into a nonlinear optimization problem (Leyffer and Munson, 2010; Steffensen and Bittner, 2014; Ralph and Smeers, 2006). They take advantage of the mathematical structure of individual modules to produce a larger problem where the existence and uniqueness properties can be studied. The disadvantage of this approach is it requires the modules to be implemented in the same programming language and requires the combined mathematical problem to be computationally tractable.

For Project BlueSky, we are designing the prototype to use both iterative and a more integrated approach, as well as a hybrid approach, where some modules are combined and then iterated with the remaining modules. The ability to combine the mathematical formulation of different modules into a larger optimization problem is a key structural enhancement versus NEMS that will improve computational performance and make the model easier to run for a variety of configurations, including sensitivity and uncertainty analyses.

### 3.2. Flexibility

As noted above, energy systems are in a period of rapid change. The next generation model must be designed flexibly so that it can be used to assess new energy and technology pathways, policies, economic conditions, and trade patterns. The boundaries and internal resolution of the modeled system also need to be flexible. For example, in some future instances, we may want to be able to model material supply chains – particularly critical minerals – and how they might impact energy technology deployment. The technological, spatial, and temporal resolutions also need to be flexible. For example, we eventually want to be able to quantify the economic impact of energy infrastructure deployments and retirements at the community scale. Expansive boundaries and maximal resolutions throughout the model would make it computationally infeasible, so the model must be designed to run easily with a customized feature set. Some aspects of specific problems could be moved outside of 'the model' but remain within a broader modeling framework, to be analyzed in detail and then brought into the simulation results.

The next generation model will incorporate features to ensure flexible operation. We maintain a strict independence between code and data: no data is hard coded in the model source code. The model source code is implemented in an abstract manner that allows it to operate on different input datasets. So, for example, changing the spatial or temporal granularity of a module would mean modifying only a configuration variable, which will trigger the model automatically adjust to accommodate the appropriate input data.

Modularity will support flexibility. Maintaining simplified versions of each module allows modelers the flexibility to configure the model by selecting the version of each module with the appropriate level of detail to answer the question at hand. It should also be possible to omit a module and instead use a set of exogenously specified inputs such as a reduced form approximation. Depending on the nature of the question at hand, it will be possible to run only select modules, run some modules with a simplified or reduced-form structure, or independently adjust module resolutions.

### 3.3. Transparency

In the context of energy models, transparency means that model documentation, code, and data are discoverable and readily understood by other modelers. Modern energy system models have unique characteristics that make transparency paramount. First, such models, aided by increased computational power, have extensive codebases that rely on large input datasets that could easily obscure the inner workings of energy models, opening them up to "black box" criticisms. Second, the results from long-term energy models are difficult to validate, so it is difficult to assess model performance based solely on comparing results with real world outcomes or intuition (Craig et al., 2002; DeCarolis et al., 2012). Third, models are instruments of inquiry that enable learning; closed models deny observers the opportunity to use the models for their own experiments and prevent rigorous peer review by other experts. Finally, model-based insights are often used by decision makers to guide policy, and thus stakeholders should be able to understand how insights that broadly affect society are derived.

NEMS includes two key features of transparency: an open-source codebase (EIA, 2024b) and highly detailed documentation (U.S. EIA, 2023). However, legacy code implementation in different programming languages and bespoke data pipelines can be difficult for external users to parse. The next generation model offers an opportunity to leverage existing EIA model transparency by using a single language with a consistent coding convention, embedded code documentation, and a well-documented data pipeline.

Making EIA's next generation model as transparent as possible conveys several advantages. First and foremost, it promotes trust among our stakeholders. Higher transparency models are also more credible because they are more understandable (Craig et al., 2002). As with all open-source projects, allowing external parties to review the code can also help identify bugs in the code and accelerate development compared to a closed process (Raymond, 1999). As noted above, the next generation model is designed to be modular and flexible. Such a structure provides a convenient test bed for modelers to experiment with new governing dynamics, computational approaches, and input data in ways that can feed back to EIA development and serve the broader modeling community in interpreting our model's results. Such an approach is only effective; however, if the model is transparent to potential users. More than three decades ago, the NRC (1992) acknowledged the benefit of making NEMS accessible to outside organizations, which offer "rich sources of information, input modules, and sectoral and geographic analyses" that would be difficult to produce internally at EIA.

### 3.4. Robustness

Uncertainty about the future is a critical consideration in long-term energy modeling. Given the complexity of the real-world system, precise predictions decades into the future are bound to be wrong (Smil, 2000). As NRC (1992) notes in their report to EIA on NEMS: "Uncertainty is inherent in the nature of models and cannot be eliminated. Nor should it be ignored." There are just too many unmodeled degrees of freedom to produce accurate forecasts in the medium-to-long term. Uncertainty is also heightened by a rapidly changing energy system driven by technology innovation and new public policy. Models represent highly simplified, self-consistent frameworks that allow modelers to explore the future decision landscape under different assumptions, and draw insights based on comparisons across multiple model runs. As experienced modelers have been imploring for decades, models should be focused on producing insights, not numbers (Huntington et al., 1982). More specifically, models can be effectively used to characterize the range of possible outcomes, examine key tradeoffs, and identify potential unforeseen outcomes associated with proposed policy.

While there are many ways to categorize uncertainty, a common framing in the context of models is a distinction between structural and parametric uncertainty (NRC, 1992; Edenhofer et al., 2006; DeCarolis 2011; Yue et al., 2018). Structural uncertainty represents the limited ability of the model's governing dynamics and equations to represent the real world. Parametric uncertainty represents the limited accuracy of the input data used in the model to align a model structure's output to real-world outcomes. In practice, many modeling studies – including EIA's *Annual Energy Outlook* – are limited to a consideration of parametric uncertainty, which is addressed by changing key input assumptions and then running different scenarios. EIA's next generation model will add significant new capabilities to quantify uncertainty. Many different methods exist to quantify future uncertainty, and the choice of method should be based on the question at hand (Yue et al., 2018).

Most methods to quantify uncertainty involve model iteration, so it is important to maintain simplified versions of each module that can be used for rapid iteration. NRC (1992) suggests the creation of reduced-form versions of each module to speed model solve time, enabling rapid model iteration and the use of probabilistic methods to assess outcomes. The choice of a simplified or detailed module for a particular exercise can be selected based on the objective of the particular analysis.

The modular structure of the next generation model also affords the opportunity to test structural uncertainty by modifying the governing equations within a particular sector. For example, modelers could assess differences in light duty vehicle deployment based on modules driven by least cost optimization and consumer choice. Such an approach allows modelers to assess how differences in model structure affect could outcomes of interest and help to develop a more complete view of future uncertainty.

Several approaches can help address parametric uncertainty. The next generation prototype includes examples of efficient methods to quantify sensitivity of model output to input parameters. One method uses an efficient way to approximate derivatives of model outputs with respect to inputs for systems of nonlinear equations using complex variables (Martins et al., 2003; Lai et al., 2005). The method is broadly applicable to any module where the equations are continuously differentiable and has been implemented in applications beyond energy modeling. A second method is directly applicable to convex optimization problems and applies approximation methods for sensitivity analysis on optimization models (Castillo et al., 2006). This method takes into account how model outputs, including active and inactive constraints, might change with changes in inputs and approximates changes in output without having to rerun the optimization. Both methods are run independently of the model and are used to calculate sensitivities after the model has produced deterministic results. This allows the model to provide parametric uncertainty in outputs without increasing solve time.

Modeling uncertainty allows us to advance the ways in which model results, the "outputs" layer in Figure 1, are shown. Often, displaying output with a heavy focus on one or two scenario runs leads to the misinterpretation that the model results are deterministic and will lead to one of a few discrete outcomes (Morgan and Keith, 2008). Modeling uncertainty not only improves our analysis capabilities but provides a clearer way to display and interpret model output. Interpreting model results is critically important and displaying them in a way that accounts for uncertainty leads to more accurate interpretation.

We intend to build new uncertainty assessment capabilities over time as particular needs arise. Monte Carlo simulation allows modelers to simultaneously vary multiple inputs and observe the distribution of outputs (Morgan and Henrion, 1990). Method as specified by Morris can be applied to rank order the input parameters that have the largest impact on output quantities of interest (Morris, 1991). For modules based on a linear programming model, modeling-to-generate alternatives (MGA) can be used to systematically search the decision space and address structural uncertainty (DeCarolis, 2011). None of these options are practically available for uncertainty assessment without a model that is also modular, flexible, and transparent.

## 4. Next Generation Model Prototype

Building a new model is challenging and requires modelers to think through all the layers shown in Figure 1. A production scale version of the next generation model with all the features described above will take several years to achieve. Model development often proceeds in a top-down fashion, starting with the governing dynamics and mathematical representation followed by code implementation. Even among open-source models, initial development is often limited within a team and only released to the public when the core model is fully functional or nearly so. As a result, the underlying code can be hard to interpret even when the model is easy to apply. We are taking a different approach to next generation model development: we want to transparently build the next generation model in stages with input from the community as development proceeds. To do so, our initial focus is on building a bare bones prototype that enables the modularity and flexibility described above.

Rather than starting at the top of Figure 1 with the selection of the most appropriate governing dynamics and mathematical formulation, we are focused on building test modules that include well-documented code and data structures and showcase core features related to modularity and flexibility. Our goal is to seek community feedback on the prototype to ensure that the modeling framework will be extensible, computationally efficient, and as transparent as possible. A well-designed framework can serve as a testbed both within EIA and the larger modeling community. Later stages of model development will include the selection of appropriate governing dynamics, mathematical formulations, and code implementation following the practices and procedures established by the first round of prototype development.

The initial prototype contains three test modules covering three sectors: electricity, hydrogen, and residential demand. The included modules are designed to share information to seek an equilibrium in prices and quantities across modules. To ensure efficient computation, the BlueSky Prototype provides two options for finding the model equilibrium: a Gauss-Seidel iteration and a hybrid optimization-iteration method. To aid the hybrid approach, we have included a generic methodology to convert non-optimization formulations into optimization formulations, using the residential module as an example. The prototype also incorporates nonlinear price responses in the residential module and endogenous technological learning with a nonlinear formulation in the electricity module. Finally, the prototype introduces approximation methods that output the sensitivity of the solution with respect to uncertain inputs for both optimization (Castillo et al., 2006) and non-optimization (Martins et al., 2003) modules.

In terms of implementation, the BlueSky Prototype draws on the advice offered by the energy modeling community (e.g., Craig et al., 2002; DeCarolis et al., 2012; Pfenninger et al., 2014). First, both model code and input data will be made publicly available under the Apache 2.0 license. Changes to model code are tracked using Git (Git, 2024), a distributed revision control system, and the codebase is publicly available through EIA's GitHub page (U.S. EIA, 2024b). Revision control allows our stakeholders to track line-by-line changes to our code over time.

Open-source code is only a prerequisite to transparency. The code and data need to be well-structured and documented. We are placing a strong emphasis on documentation, including the use of docstrings embedded in the code that can be used to auto-generate model documentation. In addition, given the complex workflows required to process and assemble model input data, we are using a workflow management tool to document and automate data processing. The prototype separates the source code from the data, allowing the flexibility of diverse data inputs without hardcoding. We recognize that transparency is a journey rather than a destination, and thus we will continue to find ways to make the next generation model easier to comprehend and use.

To minimize the barriers to entry, we also plan to use an open-source software stack to the degree possible. The model code is implemented in Python, a popular, open-source language with an extensive set of libraries available for use (Python, 2024). Pyomo is used to formulate the optimization-based modules, which includes the capability to use free or commercial solvers (Pyomo, 2024). Though we will most likely need to rely on commercial solvers for production runs due to their superior performance, the prototype is designed to work with open-source solvers, such as IPOPT (IPOPT, 2024) and HiGHS (HighsPy, 2024).

The prototype is designed with a modular structure that allows modules to be run independently and allows easy addition of new modules. The test modules feature a modular "block" structure programmed in Pyomo that allows swapping of model capabilities. We have made use of sparse indexing, mutable parameters, and shared variables in simultaneous optimization to speed up computation. The test modules include features that highlight the flexibility to adjust model size and complexity, including different temporal and spatial granularities and technological learning.

Given our focus on modularity, flexibility, and code design, the input data associated with the BlueSky Prototype is for testing purposes only. Because governing dynamics are not the focus of this prototype release, the three prototype sectoral modules borrow their governing dynamics from NEMS. In addition, the test dataset included in the prototype release are meant to be generic, i.e., the results do not pertain to any specific country, region, or timeframe. For unmodeled sectors in the prototype, such as primary energy supply, exogenous inputs are used to make the model functional.

## 5. Conclusion

When first developed in 1992, NEMS was a largely successful implementation of a vision set forth by the NRC for a broadly-scoped and modular representation of the entire U.S. energy economy. It quickly became the standard mid- and long-term model for making economy-wide projections of U.S. energy markets. Despite its success, key parts of the NRC vision for NEMS, including flexible interchange of individual sector modules, were not fully realized. Over the 30 plus years it has been in use, NEMS has accumulated numerous changes, with new modules added, older modules replaced, and significant continuous changes to models that have, notionally at least, remained "the same" as originally implemented. These factors have resulted in a model today that is patchwork quilt of modules with different governing dynamics, different coding languages, different data structures, and an unwieldy and often slow overarching solution and convergence algorithm.

EIA's BlueSky project has begun development of a model that addresses these original and accumulated concerns, as well as providing new functionality not envisioned 30 years ago. As with NEMS, each module will retain the ability to impose governing dynamics consistent with sector-specific market and technology characteristics. However, in the BlueSky prototype, each module will be coded in a single language, allowing for more efficient interoperability of modules, improved code transparency, and lower barriers-to-entry for third party users. The overall solution algorithm will allow for both iterative and global convergence approaches to improve solution efficiency. Data structures and coding practices will be well-documented and implemented consistently across modules, which will facilitate future modification or replacement of sector modules with functionally equivalent modules that may use different governing dynamics, spatiotemporal resolution, or market scope. The prototype will also contain tools to evaluate model sensitivity without the need for time-consuming and complex set-up of separate sensitivity cases or Monte-Carlo type analysis of numerous input parameters.

Once fully implemented, the modeling framework developed with this BlueSky prototype will position EIA to tackle the increasingly complex energy modeling challenges posed by both current and foreseeable technology, market, and policy developments. It will also result in a model that is not only more transparent to stakeholders such as policymakers, academics, advocates, and industry but also more accessible for outside users to modify, operate, and even extend for their own analyses.